\documentclass[aps,pre,superscriptaddress,amsmath,amssymb,preprintnumbers,showpacs,twocolumn]{revtex4-1}

\usepackage{amssymb}
\usepackage{graphicx}
\usepackage{bm}
\usepackage{color}
\usepackage{subfigure}

\usepackage{hyperref}

\begin{document}

\title{Fluctuation theorems for non-Markovian quantum processes}

\author{B. Leggio}
\affiliation{Dipartimento di Fisica e Chimica, Universit\`{a} di Palermo, Via Archirafi 36, 90123 
Palermo, Italy}
\affiliation{Physikalisches Institut, Universit\"{a}t Freiburg, 
Hermann-Herder-Stra\ss{}e 3, D-79104 Freiburg, Germany}

\author{A. Napoli}
\affiliation{Dipartimento di Fisica e Chimica, Universit\`{a} di Palermo, Via Archirafi 36, 90123 
Palermo, Italy}

\author{H.-P. Breuer}
\affiliation{Physikalisches Institut, Universit\"{a}t Freiburg, 
Hermann-Herder-Stra\ss{}e 3, D-79104 Freiburg, Germany}

\author{A. Messina}
\affiliation{Dipartimento di Fisica e Chimica, Universit\`{a} di Palermo, Via Archirafi 36, 90123 
Palermo, Italy}

\newcommand{\ket}[1]{\displaystyle{|#1\rangle}}
\newcommand{\bra}[1]{\displaystyle{\langle #1|}}

\date{\today}

\begin{abstract}
Exploiting previous results on Markovian dynamics and fluctuation theorems, we 
study the consequences of memory effects on single realizations of 
nonequilibrium processes within an open system approach. The entropy 
production along single trajectories for forward and backward processes is 
obtained with the help of a recently proposed classical-like non-Markovian 
stochastic unravelling, which is demonstrated to lead to a correction of the 
standard entropic fluctuation theorem. This correction is interpreted as
resulting from the interplay between the information extracted from the system
through measurements and the flow of information from the environment to the
open system: Due to memory effects single realizations of a dynamical 
process are no longer independent, and their correlations fundamentally affect 
the behavior of entropy fluctuations.
\end{abstract}

\pacs{03.65.Yz,05.30.Ch,05.70.Ln}

\maketitle

\section{Introduction}\label{intro}
Since many years fluctuations in far-from-equilibrium processes have been studied 
both in a classical \cite{class1,class2,class3} and in a quantum context 
\cite{quant1,quant2,quant3}. The large interest they attract is due, 
for instance, to their deep connection with equilibrium thermodynamic 
quantities characterizing a system interacting with an environment \cite{oppen, lip}, 
as well as with quantum information theory \cite{kafri}. In the quantum realm 
the consequences of such an interaction on the fluctuations of physical 
quantities of the open system has been extensively studied only under the 
Markovian approximation 
and/or in the weak coupling limit \cite{muka, broeck}. Different methods have been 
developed, allowing for the formulation of some theorems analogous to the 
well-known classical fluctuation theorems \cite{esp, def, liu}. However, an 
extension of these approaches to non-Markovian processes has only partially 
been attempted \cite{kawa} and is far from being a settled problem. Taking into 
account non-Markovian dynamics in nonequilibrium thermodynamics is, however, 
an issue of great interest, since recently memory effects have been extensively 
investigated, both from a theoretical \cite{breu1, plen, nlnm} and from and 
experimental point of view \cite{breu2}. On a more fundamental level, the 
possibility of characterizing a non-Markovian extension of thermodynamic 
concepts is very closely related to the investigation on the quantum limits of 
thermodynamics \cite{lim1, lim2}, recently attracting a great deal of attention.

In this work we extend the method developed in Ref.~\cite{muka} to the
non-Markovian regime by employing recently constructed stochastic
formulations of classical and quantum non-Markovian master equations
\cite{piilo1,piilo-pra,piilo2}. In this way we are able to account for memory effects in 
far-from-equilibrium entropic fluctuations and thus to formulate a 
non-Markovian generalization of an entropic fluctuation theorem, 
highlighting also how and why it differs from its Markovian counterpart.

The paper is structured as follows. In Sec.~\ref{masteq} we define our physical 
scenario and introduce its general dynamical description. Section \ref{markovian} 
is a brief review of the Markovian approach of \cite{muka}, slightly modified to fit 
with our goals, and explicitly emphasizes the role of the structure of 
the associated master equation. In this way the reason for which such a method 
cannot account for non-Markovian processes will become clear. 
Section \ref{nonmarkovian} is 
devoted to the introduction of the main ideas and results of this work, namely a 
generalization of \cite{muka} which exploits a stochastic unravelling of 
non-Markovian processes. In Sec.~\ref{interplay} we comment on the 
physical interpretation of the non-Markovian fluctuations described by our 
approach, and present some final remarks and conclusions.

\section{The master equation}\label{masteq}
We consider an open quantum system with free Hamiltonian $H_S$, interacting 
with an environment (free Hamiltonian $H_E$) via an interaction term $H_I$. Let 
us suppose the free Hamiltonian of the open system to show a certain time 
dependence caused by some parameters being externally modified in time. Such 
an external modification drives the system far from its initial equilibrium (or 
stationary) state and towards a final state characterized by different values of 
these control variables. During its evolution the system interacts with its 
environment, our goal being to study the effect of such an interaction on 
fluctuations of thermodynamic quantities. We employ a 
time-convolutionless (TCL) master equation \cite{shiba} for the open system 
density matrix $\rho(t)$, describing a general quantum dynamics:
\begin{equation}\label{tcl}
 \frac{d}{dt}\rho(t)=\mathcal{K}(t)\rho(t),
\end{equation}
where
\begin{eqnarray} \label{kappat}
  \!\!\!\!\! \mathcal{K}(t)\rho &=& -i\left[H(t),\rho\right] \nonumber \\
  &+& \sum_{i}\gamma_{i}(t)\bigg[A_{i}(t)
  \rho A_{i}^{\dag}(t) - \frac{1}{2}\Big\{A_{i}^{\dag}(t)A_{i}(t),\rho\Big\}\bigg] .
\end{eqnarray}
The Hamiltonian $H(t)$ in Eq.~(\ref{kappat}) describes the 
unitary part of the open system evolution, which is given by its free Hamiltonian 
$H_S(t)$ plus a renormalization term due to the interaction with the bath. The 
nonunitary part of the evolution, describing dissipation and dephasing, is 
accounted for by a set of generally time-dependent Lindblad operators $A_i(t)$ 
and corresponding relaxation rates $\gamma_i(t)$. Equation (\ref{tcl}) describes 
Markovian as well as non-Markovian processes in terms of a master equation 
which is local in time \cite{breub}.

The starting point of our analysis on nonequilibrium entropic fluctuations is the 
expression for thermodynamic ensemble quantities, and in particular for the time 
variation of the von Neumann entropy along the nonequilibrium dynamics itself. 
To this end, we write the state of our open quantum system as \cite{muka}
\begin{equation}\label{rhoeigen}
 \rho(t)=\sum_a \mu_a(t)|a(t)\rangle\langle a(t)|,
\end{equation}
where the set $\mathfrak{B}(t)=\{|a(t)\rangle\}$ is a time-dependent orthonormal 
basis of the open system Hilbert space instantaneously diagonalizing $\rho(t)$, 
and $\sum_a\mu_a(t)=1$. Using the generator of the dynamics given in 
(\ref{kappat}) and calculating the mean value of Eq.~(\ref{tcl}) for a 
state $|b(t)\rangle \in \mathfrak{B}(t)$, one easily obtains a Pauli-type master 
equation for the evolution of the populations $\mu_a(t)$,
\begin{equation} \label{dotn}
 \dot{\mu}_b(t) = \sum_a \Big(R_{ba}(t)\mu_a(t) - R_{ab}(t)\mu_b(t) \Big),
\end{equation}
where the total instantaneous transition rate $R_{ba}(t)$ between two states 
$|a(t)\rangle$ and $|b(t)\rangle$ belonging to $\mathfrak{B}(t)$ is defined as
\begin{equation}\label{totrates}
 R_{ba}(t)=\sum_i \gamma_i(t) |\langle b(t)|A_i(t)|a(t)\rangle |^2.
\end{equation}
These transition rates will turn out to be crucial in the expression for all 
thermodynamic quantities of interest. Moreover, it is their time behavior which we 
employ to characterize the occurrence of memory effects during the process: As 
will be discussed in more detail in Sec.~\ref{nonmarkovian}, the process
described by the Pauli master equation (\ref{dotn}) is defined to be 
non-Markovian if and only if at least one of the transition rates $R_{ba}(t)$ 
between two particular instantaneous eigenstates of $\rho(t)$ temporarily 
becomes negative.

\section{The Markovian case}\label{markovian}
In the case of a purely (and possibly time-dependent) Markovian dynamics, the 
rates given by Eq.~(\ref{totrates}) never become negative. In what follows we will 
use indices $a$ and $b$ to label vectors in the instantaneous eigenbasis of 
$\rho(t)$ (recall Eq.~(\ref{rhoeigen})), and an index $i$ to label the possible decay 
channels described by the set of Lindblad operators $A_i(t)$ in (\ref{kappat}). We 
will furthermore, for the sake of brevity, sometimes suppress the time 
arguments of the instantaneous eigenvectors and eigenvalues of $\rho(t)$.

\subsection{Entropies}
Evaluating the von Neumann entropy in the instantaneous 
eigenbasis of $\rho(t)$, we obtain its time derivative $\dot{S}$ in the form
\begin{equation} \label{dotsv}
 \dot{S}(t)=-\sum_b\dot{\mu}_b(t)\ln \mu_b(t).
\end{equation}
Using Eq.~(\ref{dotn}) in Eq.~(\ref{dotsv}) we find
\begin{equation} \label{dotsvm}
 \dot{S}(t) = -\sum_{a,b}\mu_a(t)R_{ba}(t)\ln\frac{\mu_b(t)}{\mu_a(t)}.
\end{equation}
A similar expression was derived in Ref.~\cite{muka}, but the use of a TCL master 
equation in our approach allows us to express transition rates explicitly in terms of 
Lindblad operators. This clarifies the physical framework we are working in and, as 
will become evident in Sec.~\ref{nonmarkovian}, explicitly shows where and how 
memory effects come into play in the case of non-Markovian dynamics.

Having at our disposal the expression for the time derivative of entropy, and 
following usual prescriptions of nonequilibrium thermodynamics \cite{schnak,leb}, 
we can write Eq.~(\ref{dotsvm}) as a sum of two different contributions as 
$\dot{S}(t)=\dot{S}_e(t)+\dot{S}_i(t)$, having defined
\begin{eqnarray}
 \dot{S}_e(t) &=& -\sum_{a,b}\mu_a(t)R_{ba}(t)\ln\frac{R_{ba}(t)}{R_{ab}(t)}, 
 \label{entflux} \\
 \dot{S}_i(t) &=& \sum_{a,b}
 \mu_a(t)R_{ba}(t)\ln\frac{\mu_a(t)R_{ba}(t)}{\mu_b(t)R_{ab}(t)}
 \label{entprod}
\end{eqnarray}
as, respectively, the entropy flux between system and environment and the total 
entropy production. Equations (\ref{dotsvm}), (\ref{entflux}) and ({\ref{entprod}) 
describe the time dependence of entropy due to the ensemble dynamics described 
by a TCL master equation. The irreversibility of the process is characterized by a 
nonzero rate of entropy production $\dot{S}_i(t)$ inside the system which, 
furthermore, in the case of a Markovian dynamics never becomes negative. It is 
worth stressing that the definition (\ref{entprod}) for the entropy production 
coincides with the negative time derivative of the relative entropy of $\rho(t)$ and 
the stationary state $\rho_{\mathrm{stat}}$ of the dynamics, provided 
the latter exists and detailed balance holds.

\subsection{Fluctuations}\label{markfluct}
The above defined quantities characterize the physics of the open quantum 
system on an ensemble level. Such a picture, while allowing us to derive suitable 
expressions for many quantities of interest, lacks however a characterization of 
single nonequilibrium processes and, in particular, of their intrinsically fluctuating 
physical quantities. The goal of this section is to obtain a fluctuation theorem for 
the entropy production along single realizations of the ensemble dynamics. By 
definition, a fluctuation theorem for a quantity $Q$ characterizing a thermodynamic 
system is an expression for the ratio of the probability of such a quantity having 
the value $q$ along a particular nonequilibrium process, and the probability of the 
same quantity having a value $-q$ along the backward realization of the same 
process. In order to describe these fluctuations we employ the master equation
(\ref{dotn}) which in the present case only involves positive transition
rates and can thus be regarded as a differential Chapman-Kolmogorov equation 
for a classical, Markovian stochastic jump process with, in general, 
time-dependent rates $R_{ba}(t)$. Let us consider a particular, yet generic
realization of this process,
\begin{equation} \label{forw}
 |a_0(t_0)\rangle \to |a_1(t_1)\rangle \to \cdots \to |a_N(t_N)\rangle,
\end{equation}
consisting of $N$ jumps at times $t_1,t_2,\ldots,t_N$ between well defined states 
belonging to the instantaneous eigenbases of $\rho(t)$. According to the master
equation (\ref{dotn}) the probability of the trajectory (\ref{forw}) can be written as
\begin{eqnarray} \label{pfm}
 p_f &=&  \mu_{a_0}(t_0)\prod_{j=0}^{N-1}
 e^{-\int_{t_j}^{t_{j+1}}\mathrm{d}\tau\sum_{b}R_{ba_{j}}(\tau)} \nonumber \\
 && \qquad \; \times \prod_{j=0}^{N-1}R_{a_{j+1}a_j}(t_{j+1})\mathrm{d}t_{j+1},
\end{eqnarray}
where the first factor gives the probability for the system to start its 
trajectory in the state $|a_0\rangle$, the second factor gives the probability
that there are no jumps between the times $t_j$ and $t_{j+1}$, and the third factor
represents the probability of having $N$ jumps within infinitesimal time intervals
$\mathrm{d}t_j$ around $t_j$ between the states of the trajectory (\ref{forw}).

It is important to emphasize that the stochastic description given by
Eq.~(\ref{pfm}) is based on the Pauli-type master equation (\ref{dotn}) and
thus correspond to the standard stochastic unraveling of a classical Markovian 
master equation \cite{gill}, in which the probability for a transition from state 
$|a(t)\rangle$ to state $|b(t)\rangle$ during the time interval $\mathrm{d}t$ is 
determined by $R_{ba}(t)\mathrm{d}t$ with the rate given by Eq.~(\ref{totrates}). 
Thus we follow here the interpretation proposed in \cite{muka} to identify
the fluctuations of single nonequilibrium processes with those described
by the evolution equation (\ref{dotn}) for the populations of the density matrix.
This interpretation and the underlying physical picture has to be carefully
distinguished from the interpretation of the stochastic wave function methods 
(see, e.g., Ref.~\cite{mcwf} and references therein) for the open system dynamics 
given by a quantum master equation of the form of Eq.~(\ref{tcl}) in terms of a 
continuous measurement of the environment.

The backward process corresponding to (\ref{forw}) is described by the trajectory
\begin{equation}\label{backw}
 |a_N(t_N)\rangle \to |a_{N-1}(t_{N-1})\rangle \to \cdots \to |a_0(t_0)\rangle.
\end{equation}
The probability $p_b$ for such a backward process is defined analogously to what 
has been done for the forward one, taking into account the jumps in the sequence 
given in Eq.~(\ref{backw}) such that the conditioned non-jump evolution probability 
is then the same for forward and for backward processes, while the jump rates are 
reversed,
\begin{eqnarray} \label{pbm}
 p_b &=&  \mu_{a_N}(t_N)\prod_{j=0}^{N-1}
 e^{-\int_{t_j}^{t_{j+1}}\mathrm{d}\tau\sum_{b}R_{ba_{j}}(\tau)} \nonumber \\
 && \qquad \;\;\; \times \prod_{j=0}^{N-1}R_{a_ja_{j+1}}(t_{j+1})\mathrm{d}t_{j+1}.
\end{eqnarray}
Thus, we find that the logarithm of the ratio of the two probabilities takes the form
\begin{equation}\label{rmt}
 \ln\frac{p_f}{p_b}=\ln\frac{\mu_{a_0}(t_0)}{\mu_{a_N}(t_N)}+\sum_{j=0}^{N-1}
 \ln\frac{R_{a_{j+1}a_j}(t_{j+1})}{R_{a_ja_{j+1}}(t_{j+1})}.
\end{equation}
The first term
\begin{equation} 
 \Delta s = \ln\frac{\mu_{a_0}(t_0)}{\mu_{a_N}(t_N)} \label{del-ent-vn}
\end{equation}
represents the change of the von Neumann entropy, while
\begin{equation}
 \Delta s_e = -\sum_{j=0}^{N-1}
 \ln\frac{R_{a_{j+1}a_j}(t_{j+1})}{R_{a_ja_{j+1}}(t_{j+1})} \label{del-ent-flux}
\end{equation}
yields the entropy flux integrated along a single trajectory. The average over all 
possible trajectories leads to the expressions (\ref{dotsvm}) and 
(\ref{entflux}), respectively. Defining
\begin{equation}
 \sigma=\Delta s-\Delta s_e
\end{equation}
as the total entropy production along a single trajectory, we thus obtain an entropic
fluctuation theorem for Markovian dynamics,
\begin{equation} \label{fluctm}
 \frac{p_f(\sigma)}{p_b(-\sigma)} = e^{\sigma},
\end{equation}
from which the quantum analog of Crooks theorem \cite{espomuka} and the 
quantum Jarzynski equality \cite{hanggi} directly follow.

\section{The non-Markovian case}\label{nonmarkovian}

What happens if we introduce memory effects into the dynamics? The 
generator (\ref{kappat}) of the TCL master equation describing a non-Markovian 
time evolution keeps the same structure as before, but the decay rates 
$\gamma_i(t)$ can temporarily become negative. This allows us to describe in full 
generality any non-Markovian quantum process, the TCL formulation being very 
general, requiring only the map to be invertible and differentiable with respect to 
time \cite{stel}. We are going to study the thermodynamic consequences of this 
behavior. An attempt to take into account negative decay rates in the formulation 
of a fluctuation theorem has already been performed in \cite{kawa} where, 
however, a clear formulation of a non-Markovian fluctuation theorem was not 
given. In particular the approach of \cite{kawa} works well as long as the transition 
rates $R_{ba}(t)$ stay positive, which however is the characterization we gave of a 
Markovian process. A study of nonequilibrium fluctuations, then, has not been 
performed yet for the class of processes which we define as non-Markovian. What 
we are going to develop is, on the contrary, a formulation valid in any case, also 
including our definition of non-Markovianity.

\subsection{Renormalized entropies}\label{nonmarkent}

Analogously to what has been done in the Markovian case, we start by 
analyzing the entropic ensemble behavior of the open quantum system.
The equation for the time derivative of von Neumann entropy, Eq.~(\ref{dotsvm}),
is of course the same, but now the time evolution of populations is affected by 
memory effects. To see where exactly these effects come into play, let us closely 
analyze Eq.~(\ref{dotn}) which follows from any TCL master equation, either 
Markovian or not. Such an expression for $\dot{\mu}_b(t)$ depends on the 
transition rates $R_{ba}(t)$ defined in (\ref{totrates}). Notice however that each 
term in the sum on the right-hand side of (\ref{totrates}) is proportional to a decay 
rate $\gamma_i(t)$ which, in the non-Markovian case, may cause the whole sum 
to temporarily become negative. To deal with this, let us rewrite the total transition 
rates as
\begin{equation} \label{totratesnm}
 R_{ba}(t) = R_{ba}^{\mathrm{M}}(t)-R_{ba}^{\mathrm{NM}}(t),
\end{equation}
where the Markovian contribution $R_{ba}^{\mathrm{M}}(t)$ and the 
non-Markovian contribution $R_{ba}^{\mathrm{NM}}(t)$ are defined by
\begin{eqnarray}
 R_{ba}^{\mathrm{M}}(t) &=& \frac{1}{2} \big[ |R_{ba}(t)| + R_{ba}(t) \big], \\
 R_{ba}^{\mathrm{NM}}(t) &=& \frac{1}{2} \big[ |R_{ba}(t)| - R_{ba}(t) \big] . 
\end{eqnarray}
With these definitions, Eq.~(\ref{dotsvm}) can be rewritten as
\begin{equation} \label{dotsvnm}
 \dot{S}(t) = -\sum_{a,b}\mu_a(t)
 \Big( R_{ba}^{\mathrm{M}}(t)-R_{ba}^{\mathrm{NM}}(t) \Big)
 \ln\frac{\mu_b(t)}{\mu_a(t)}.
\end{equation}
The occurrence of memory effects has the consequence of reducing the rate of 
change of entropy of the system during certain intervals of time. It is possible, as 
shown, to separate the non-Markovian contribution to the change of von Neumann 
entropy and to prove that such a contribution always counteracts the Markovian 
one.

The next step is now to define entropy flux and entropy production for the 
analyzed process. In particular it is interesting to investigate the possibility of 
singling out non-Markovian contributions in these two quantities, as done for the 
von Neumann entropy itself. We then proceed as before, writing
\begin{equation}\label{split}
 \dot{S}(t)=\dot{\mathfrak{S}}_e(t)+\dot{\mathfrak{S}}_i(t),
\end{equation}
where formally both $\dot{\mathfrak{S}}_e(t)$ and $\dot{\mathfrak{S}}_i(t)$ are the 
entropy flux and production for the ensemble dynamics, and have the same 
expression as in the Markovian case, but they are fundamentally different because 
of the new structure of the transition rates $R_{ba}(t)$. Indeed, we have
\begin{eqnarray} 
 \dot{\mathfrak{S}}_e(t) &=& -\sum_{a,b}\mu_a(t)
 \Big( R_{ba}^{\mathrm{M}}(t)-R_{ba}^{\mathrm{NM}}(t) \Big) 
 \label{complexent-flow} \nonumber \\
 && \qquad \times\ln\frac{ R_{ba}^{\mathrm{M}}(t)-R_{ba}^{\mathrm{NM}}(t)}{ 
 R_{ab}^{\mathrm{M}}(t)-R_{ab}^{\mathrm{NM}}(t)}, \\
 \dot{\mathfrak{S}}_i(t) &=& \sum_{a,b}\mu_a(t)\Big( R_{ba}^{\mathrm{M}}(t)-R_
 {ba}^{\mathrm{NM}}(t) \Big) \nonumber \\
 && \qquad \times\ln\frac{ \mu_a(t) \Big(R_{ba}^{\mathrm{M}}(t)
 -R_{ba}^{\mathrm{NM}}(t)\Big)}{ \mu_b(t)\Big(R_{ab}^{\mathrm{M}}(t)
 -R_{ab}^{\mathrm{NM}}(t)\Big)}. \label{complexent-prod}
\end{eqnarray}
However these quantities, as already highlighted in \cite{kawa}, involve 
logarithms of not necessarily positive terms and may become temporarily 
ill-defined (although both the von Neumann entropy and its time derivative are of 
course always mathematically well-defined). Moreover, it is not 
possible to clearly isolate a Markovian and a non-Markovian contribution to these 
quantities, since the arguments of both logarithms involve Markovian and 
non-Markovian effects in a non-factorizable way.

How is it then possible to speak about ensemble entropy production along 
non-Markovian dynamics? The appearance of negative rates is, by our definition, 
the characterization of non-Markovianity and, moreover, it is known that 
these rates cannot all be negative at the same time, which means there will 
always be at least one negative ratio involved in the definition of entropy flux and 
production. To overcome this problem, let us consider again Eq.~(\ref{split}). It 
simply amounts to writing the time derivative of von Neumann entropy as a sum of 
two contributions, each of which may involve the logarithm of a negative number. 
Notice however that, apart from the factor $\mu_a/\mu_b$ (which can 
be written as an additional logarithmic term in the sum), the arguments of each 
logarithm in $\dot{\mathfrak{S}}_e(t)$ and $\dot{\mathfrak{S}}_i(t)$ are the same. 
If we replace all negative ratios $R_{ba}/R_{ab}$ under the 
logarithms by their moduli $|R_{ba}/R_{ab}|$, the decomposition
of $\dot{S}(t)$ into the two contributions still holds, since 
$\ln|R_{ba}/R_{ab}|$ is added and subtracted in the sum.
It is thus natural to define entropy flux and production for the non-Markovian 
ensemble dynamics as
\begin{eqnarray} 
 \dot{S}_e(t) &=& -\sum_{a,b}\mu_a(t)\Big( R_{ba}^{\mathrm{M}}(t)
 -R_{ba}^{\mathrm{NM}}(t) \Big) \nonumber \\
 && \qquad \times\ln\frac{ \Big|R_{ba}^{\mathrm{M}}(t)-R_{ba}^{\mathrm{NM}}(t)
 \Big|}{\Big|R_{ab}^{\mathrm{M}}(t)-R_{ab}^{\mathrm{NM}}(t)\Big|}, 
 \label{entflux-nm} \\
 \dot{S}_i(t) &=& \sum_{a,b}\mu_a(t)\Big( R_{ba}^{\mathrm{M}}(t)-R_{ba}^{\mathrm
 {NM}}(t) \Big) \nonumber \\
 && \qquad \times \ln\frac{ \mu_a(t) \Big|R_{ba}^{\mathrm{M}}(t)
 -R_{ba}^{\mathrm{NM}}(t)
 \Big|}{ \mu_b(t)\Big|R_{ab}^{\mathrm{M}}(t)-R_{ab}^{\mathrm{NM}}(t)\Big|}.
 \label{entprod-nm}
\end{eqnarray}
These definitions coincide with the ones given in Eqs.~(\ref{complexent-flow}) 
and (\ref{complexent-prod}) when the latter are real quantities, and extend them to 
general non-Markovian ensemble dynamics. Equations (\ref{dotsvnm}) and 
(\ref{entprod-nm}) clearly show how memory effects manifest themselves in 
backflows of information from the environment to the system as now $\dot{S}_i(t)$ 
can become negative.

There is however a second problem which is closely connected to the very 
definition of a fluctuation theorem: As in the Pauli master equation (\ref{dotn}) 
some rates are negative, it is not possible to give it a pure state single trajectory 
description. Nevertheless, it is possible to proceed along a slightly different path 
which will indeed lead to a theorem for out-of-equilibrium fluctuations. To this end,
consider Eq.~(\ref{dotsvnm}) which is exact and directly stems 
from the TCL master equation we started from. It can be rewritten as follows,
\begin{eqnarray} \label{svnmr}
 \dot{S}(t) &=& -\sum_{a,b}\mu_a(t)\Big( R_{ba}^{\mathrm{M}}(t)+\frac{\mu_b(t)} 
 {\mu_a(t)}R_{ab}^{\mathrm{NM}}(t) \Big)\ln\frac{\mu_b(t)}{\mu_a(t)} \nonumber \\
 &=& -\sum_{a,b}\mu_a(t)T_{ba}(t)\ln\frac{\mu_b(t)}{\mu_a(t)}.
\end{eqnarray}
where we have introduced positive {\textit{renormalized}} transition rates 
\begin{equation} \label{T_BA}
 T_{ba}(t) = R_{ba}^{\mathrm{M}}(t)
 + \frac{\mu_b(t)}{\mu_a(t)}R_{ab}^{\mathrm{NM}}(t).
\end{equation}
Since positive transition rates characterize Markovian dynamics we can exploit 
these renormalized rates to define effective Markovian-like flux 
and production for the open system entropy by means of
\begin{eqnarray} 
 \dot{S}_e^{r}(t) &=& -\sum_{a,b}\mu_aT_{ba}\ln\frac{T_{ba}}{T_{ab}}, 
 \label{entflux-r} \\
 \dot{S}_i^{r}(t) &=& \sum_{a,b}\mu_aT_{ba}\ln\frac{\mu_aT_{ba}}{\mu_bT_{ab}}.
 \label{entprod-r}
\end{eqnarray}
These two quantities are always well defined from a mathematical point of view, 
and their sum just gives back Eq.~(\ref{dotsvnm}). The microscopic motivation for 
these definitions will be given in Sec.~\ref{nonmarkfluct}. For now let us just look at 
Eqs.~(\ref{entflux-r}) and (\ref{entprod-r}) as effective quantities, which on one 
hand solve the problem of negative arguments of the logarithms, and on the 
other hand reduce to Eqs.~(\ref{entflux}) and (\ref{entprod}) in the limit of a 
Markovian dynamics. We can consider these quantities as the Markovian part of 
the expressions (\ref{entflux-nm}) and (\ref{entprod-nm}). The remaining part, 
which can not be effectively described as Markovian and which is thus irreducibly 
non-Markovian, is given by the difference between the quantities (\ref{entflux-nm})
and (\ref{entprod-nm}), obtained as a direct extension of the Markovian ones, and 
the renormalized quantities (\ref{entflux-r}) and (\ref{entprod-r}),
\begin{eqnarray} \label{irr}
 \dot{S}_X(t) &\equiv& \dot{S}_e(t)-\dot{S}_e^{r}(t)=\dot{S}_i^{r}(t)-\dot{S}_i(t)
 \nonumber \\
 &=& \sum_{a,b} \mu_aT_{ba}\ln\frac{T_{ba}|R_{ab}|}{|R_{ba}|T_{ab}}.
\end{eqnarray}
We note that the quantity $\dot{S}_X(t)$ is zero if all the renormalized transition 
rates are equal to the original ones, i.e., if all transition rates $R_{ba}(t)$ are 
positive and there are no signatures of memory effects.

\subsection{Non-Markovian fluctuations}\label{nonmarkfluct}

In order to formulate non-Markovian fluctuations of physical quantities along
single trajectories we first observe that the master equation (\ref{dotn}) can be
rewritten in terms of the renormalized transition rates (\ref{T_BA}) as
\begin{equation} \label{pauli-nm}
 \dot{\mu}_b(t) = \sum_a \Big(T_{ba}(t)\mu_a(t) - T_{ab}(t) \mu_b(t) \Big).
\end{equation}
This form of the master equation is strongly suggested by Eq.~(\ref{svnmr}) for the 
time derivative of the von Neumann entropy and the corresponding decomposition 
into renormalized entropy flux and production given by Eqs.~(\ref{entflux-r}) and 
(\ref{entprod-r}). Note that Eq.~(\ref{pauli-nm}) holds for both Markovian and
non-Markovian processes, and that the renormalized rates $T_{ba}(t)$ are
always positive by construction. As discussed in \cite{piilo2}, the form
(\ref{pauli-nm}) of the master equation emerges if one interprets a negative rate 
$R_{ba}$ for a transition from state $a$ to state $b$ as an effective positive rate 
for the reversed transition which is given by 
$T_{ba}=\frac{\mu_b}{\mu_a}|R_{ab}|$ according to Eq.~(\ref{T_BA}). Thus,
we suggest employing the master equation (\ref{pauli-nm}) for the description
of fluctuations along single realizations of nonequilibrium processes. It should
be noted however that in the non-Markovian case the transition rates 
$T_{ba}(t)$ depend on the occupation probabilities and that, therefore, different
trajectories are no longer independent which expresses the presence of
memory effects \cite{piilo-pra,epl-nm-jumps}.

Considering again a particular forward trajectory given by (\ref{forw}) we then
find the corresponding probability
\begin{eqnarray} \label{pf}
 p_f &=& \mu_{a_0}(t_0)\prod_{j=0}^{N-1}e^{-\int_{t_j}^{t_{j+1}}\mathrm{d}\tau
 \sum_{b} T_{ba_j}(\tau)} \nonumber \\
 && \qquad \; \times\prod_{j=0}^{N-1} T_{a_{j+1}a_j}(t_{j+1}) \mathrm{d}t_{j+1},
\end{eqnarray}
simply by replacing the original transition rates by the renormalized ones.
Correspondingly, the probability for the backward trajectory is given by 
\begin{eqnarray} \label{pb}
 p_b &=& \mu_{a_N}(t_N)\prod_{j=0}^{N-1}e^{-\int_{t_j}^{t_{j+1}}\mathrm{d}\tau
 \sum_{b} T_{ba_j}(\tau)} \nonumber \\
 && \qquad \;\;\; \times\prod_{j=0}^{N-1} T_{a_ja_{j+1}}(t_{j+1}) \mathrm{d}t_{j+1},
\end{eqnarray}
and we obain for the logarithm of the ratio of forward and backward probability
\begin{equation} \label{rrnm}
 \ln\frac{p_f}{p_b}=\ln\frac{\mu_{a_0}(t_0)}{\mu_{a_N}(t_N)}+\sum_{j=0}^{N-1}\ln 
 \frac{T_{a_{j+1}a_j}(t_{j+1})}{T_{a_ja_{j+1}}(t_{j+1})}.
\end{equation}
The first term on the right-hand side is again equal to the change of the
von Neumann entropy $\Delta s$ along the trajectory (see 
Eq.~(\ref{del-ent-vn})). In analogy to the Markovian case, the second term
represents the negative of the entropy flux integrated along the trajectory, 
i.~e. we have (compare with Eq.~(\ref{del-ent-flux}))
\begin{equation} \label{sigma-renormalized} 
 \Delta s^r_e = -\sum_{j=0}^{N-1}
 \ln\frac{T_{a_{j+1}a_j}(t_{j+1})}{T_{a_ja_{j+1}}(t_{j+1})}.
\end{equation}
Defining the renormalized single trajectory entropy production 
\begin{equation} \label{sigma-r}
 \sigma_r = \Delta s - \Delta s^r_e
\end{equation}
we immediately obtain from Eq.~(\ref{rrnm}):
\begin{equation}\label{fluct1}
 \frac{p_f(\sigma_r)}{p_b(-\sigma_r)}=e^{\sigma_r}.
\end{equation}
Thus, we have found a fluctuation theorem for non-Markovian processes
which is formally identical to the one obtained for Markovian dynamics, see
Eq.~(\ref{fluctm}). However, in the non-Markovian case the fluctuation theorem 
holds for the renormalized entropy production $\sigma_r$ which can be written as
\begin{equation} \label{del-sigma-r}
 \sigma_r = \Delta s - \Delta s_e + \Delta s_X = \sigma + \Delta s_X,
\end{equation}
where $\Delta s_X$ is the single trajectory contribution to the time integral
over $\dot{S}_X(t)$ (see Eq.~(\ref{irr})). In the Markovian case $\Delta s_X$ 
vanishes and Eq.~(\ref{fluct1}) reduces to Eq.~(\ref{fluctm}).
This means that in general only a part of the entropy production, namely the one 
originating from the fluctuations described by the form (\ref{pauli-nm}) of the 
master equation can be described in terms of a fluctuation theorem. 
The fluctuation theorem for non-Markovian processes is thus fundamentally 
different from its Markovian counterpart, as it describes fluctuations of the entropy 
production of single processes, which are not the single trajectory 
contribution to the ensemble entropy production (\ref{entprod-nm}).

\begin{center}
\begin{figure}
\includegraphics[width=0.4\textwidth]{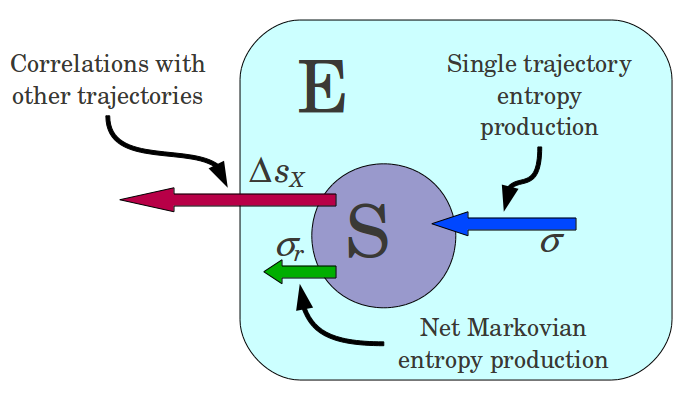}
\caption{(Color online) Schematic view of the contributions to the entropy
production along single nonequilibrium trajectories according to 
Eq.~(\ref{del-sigma-r}). The net Markovian information flux $\sigma_r$, which
obeys the fluctuation theorem (\ref{fluct1}), is equal 
to the sum of the single trajectory entropy production $\sigma$ and the 
non-Markovian contribution $\Delta s_X$ due to correlations between trajectories.}
\label{scheme}
\end{figure}
\end{center}

To interpret the result (\ref{fluct1}) in physical terms we first note that 
within our approach fluctuations are described by the stochastic unraveling of the 
master equation (\ref{pauli-nm}). However, as has been emphasized already, due 
to the presence of memory effects single trajectories are not independent of each 
other or, in other words, they are correlated. Due to these correlations a single 
realization of a nonequilibrium non-Markovian process thus yields on average 
more information than just the one described by its associated entropy production.
This is why the measured renormalized entropy production, obtained in this 
work as $\sigma_r$ in Eq.~(\ref{del-sigma-r}), is given by the usual single 
trajectory entropy production $\sigma$ {\textit{plus}} an additional term which 
alters the usual Markovian form of the fluctuation theorem. This term, given by 
$\Delta s_X$ in Eq.~(\ref{del-sigma-r}), represents the additional information 
extracted from the system 
originating from the correlations between single trajectories. As we have 
demonstrated these contributions combine such that their sum obeys the 
fluctuation theorem (\ref{fluct1}), completely analogous to the classical and the 
quantum Markovian one (see Fig.~\ref{scheme}).

\section{Conclusions}\label{interplay}

The analysis of Sec.~\ref{nonmarkovian} shows that, when taking 
into account memory effects in a quantum thermodynamics context, there is a 
close link between fluctuations in entropy production along single realizations of 
nonequilibrium processes and the existence of an irreducibly non-Markovian 
entropic contribution in the ensemble dynamics. Any time the 
dynamics shows signatures of memory effects, a quantum fluctuation theorem has 
to take into account the full information contribution of the dynamics, which is no 
longer given only by the ensemble entropy production. 
More precisely, when a single trajectory is taken into account, 
memory effects produce an additional term to the information an external 
observer can reveal by measurements. 
Such an additional measured information contribution originates from the 
existence of correlations between single trajectories, which in turn is a 
consequence of memory 
effects. This can be clearly seen from Eq.~(\ref{pauli-nm}), as the differential 
equations for the time evolution of populations are no longer linear due to the 
structure of renormalized rates, Eq.~(\ref{T_BA}).
Performing a measurement of entropy production along a single trajectory, then, 
means also extracting information about any other possible trajectory, and we 
have demonstrated that this extracted 
information leads to a net entropy production which effectively behaves 
as a Markovian one. It is this effective Markovian entropy production along single 
trajectories whose fluctuations can be described by means of a fluctuation 
theorem of the usual form. The effective Markovian entropy production is obtained 
as the sum of the single trajectory contribution to ensemble entropy production 
and the information on trajectories correlations extracted by measurements. Quite 
remarkably, these two terms combine together in such a way that their sum 
behaves according to a very simple fluctuation law.

Thus, due to non-Markovian features the measurement of fluctuating quantities 
affects the fluctuations themselves. In a sense, the fluctuations we reveal in our 
approach do not describe only single trajectory properties but supply information 
on the whole set of possible realizations of a thermodynamic process and on their 
mutual dependence. The renormalization of rates performed in 
Eqs.~(\ref{entflux-r}) and (\ref{entprod-r}) is, indeed, nothing but the mathematical 
counterpart of the scheme depicted in Fig.~\ref{scheme}. We expect, however,
that the fluctuation theorem (\ref{fluct1}) might undergo more deep modifications 
if one considers other measurement schemes in order to characterize single
trajectories, such as stochastic wave function unravellings expressing a 
continuous monitoring of the environment. This point could be an important and
fruitful subject of future works in this field.

\end{document}